\documentclass[aps,twocolumn,prb]{revtex4-1}

\usepackage{txfonts}
\usepackage{hyperref}
\usepackage{graphicx}
\usepackage{color}
\usepackage{lineno}

\hypersetup{colorlinks,
linkcolor=blue,
filecolor=blue,
urlcolor=blue,
citecolor=red}

\begin{document}
\title{Anisotropic Thermoreflectance Thermometry: \\
A contactless frequency-domain approach to study anisotropic thermal transport}

\author{L. A. P\'erez$\,^1$}
\thanks{Authors contributed equally to this work}
\author{K. Xu$\,^1$}
\thanks{Authors contributed equally to this work}
\author{M. R. Wagner$\,^2$}
\author{B. D\"orling$\,^1$}
\author{A. Perevedentsev$\,^1$}
\author{A. R. Go\~ni$\,^{1,3}$}
\author{M. Campoy-Quiles$\,^1$}
\author{M. I. Alonso$\,^1$}
\author{J. S. Reparaz$\,^1$}
\email[Corresponding author:]{jsreparaz@icmab.es}

\address{$\,^1$Institut de Ci\`encia de Materials de Barcelona, ICMAB-CSIC, Campus UAB, 08193 Bellaterra, Spain.}
\address{$\,^2$Institut f\"ur Festk\"orperphysik, Technische Universit\"at Berlin, Hardenbergstr. 36, 10623 Berlin, Germany}
\address{$\,^3$ICREA, Passeig Llu\'{\i}s Companys 23, 08010 Barcelona, Spain}


\begin{abstract}
We developed a novel contactless frequency-domain approach to study thermal transport, which is particularly convenient when thermally anisotropic materials are considered.
The method is based on a similar line-shaped heater geometry as used in the 3-omega method, however, keeping all the technical advantages offered by non-contact methodologies. 
The present method is especially suitable to determine all the elements of the thermal conductivity tensor, which is experimentally achieved by simply rotating the sample with respect to the line-shaped optical heater.
We provide the mathematical solution of the heat equation for the cases of anisotropic substrates, multilayers, as well as thin films. This methodology allows an accurate determination of the thermal conductivity, and does not require complex modeling or intensive computational efforts to process the experimental data, i.e., the thermal conductivity is obtained through a simple linear fit (``slope method''), in a similar fashion as in the 3-omega method. We demonstrate the potential of this approach by studying isotropic and anisotropic materials in a wide range of thermal conductivities. In particular, we have studied the following inorganic and organic systems: (i) glass, Si, and Ge substrates (isotropic), (ii) $\beta-$Ga$_2$O$_3$, and a Kapton substrate (anisotropic) and, (iii) a 285 nm SiO$_2$/Si thin film. The accuracy in the determination of the thermal conductivity is estimated at $\approx$ 5\%, whereas the best temperature resolution is $\Delta T \approx 3$ mK.

\end{abstract}


\maketitle

\section{Introduction}

The study of thermal anisotropy in solids has recently drawn considerable scientific and technological attention in different fields.\cite{Renteria2015,Jang2015a,Lee2015b,Li2017,Romano2017,Jiang2018b,Kubis2020,Sun2019,Ly2021}
The ability to control heat transport through intrinsic or purposely engineered thermal anisotropy, has the potential to open new routes to develop novel concepts towards smart heat manipulation.\cite{Zhu2014,Li2012h,Maldovan2013b,Li2021}
In thermally anisotropic materials, thermal anisotropy is evidenced by the different elements of the thermal conductivity tensor ($\kappa_{ij})$, leading to the tensorial expression of Fourier's law, $q_i=\kappa_{ij}(\partial T/\partial x^j$), where $q_i$ are each of the components of the vectorial heat flux, $T$ is the temperature, and $x^j$ is the spatial coordinate. The development of novel experimental methodologies  to study anisotropic thermal transport has recently become a relevant research objective. A considerable number of experimental techniques and methodologies
\cite{Tong2006,Ramu2012,Mishra2015a,Schmidt2008,Feser2012a,Feser2014,Rodin2017,Jiang2017,Jiang2018, Li2018,Rahman2018,Yuan2019,Qian2020,Tang2021,Ly2021} based on variations of the 3-omega method,\cite{Cahill1990} time-domain thermoreflectance,\cite{Cahill2004} and frequency-domain thermoreflectance\cite{Kwon2017} have been developed for this purpose,  demonstrating their capability to obtain the components of $\kappa_{ij}$. The main differences between these approaches are the dimensionality of the heat source (line or spot), and their contact or contactless fashion (electrical resistor or focused optical spot). 
The techniques developed in Refs. [\onlinecite{Ramu2012}-\onlinecite{Mishra2015a}] are based on the  geometry used in the 3-omega method, i.e., a long and narrow line-shaped heat source. This geometry provides as main advantage, simultaneous thermal sensitivity to only two crystallographic directions, which arises from the 1-dimensional (1D) geometry of the heat source. In other words, the temperature distribution in the direction along the heat source (long narrow line) is uniform, hence, no heat flow occurs in this direction, considerably simplifying the data analysis process. The main drawbacks of the approaches given in Refs. [\onlinecite{Ramu2012,Mishra2015a}] are: (i) that  electrical contacts must be deposited and contacted, and (ii) the technical difficulties in studying electrical conductors, since the transducer must be electrically insulated from the sample by an intermediate insulating layer to avoid leakage of the electrical current used to heat the resistor.
On the other hand, contactless techniques such as those reported in Refs. [\onlinecite{Schmidt2008,Feser2012a, Feser2014, Rodin2017,Jiang2017, Jiang2018, Li2018, Rahman2018,Yuan2019,Qian2020,Tang2021,Ly2021}] are based on small Gaussian or ellipse-shaped focused spots, i.e., sensitive to all crystallographic directions simultaneously. Although this is not an intrinsic impediment to obtain $\kappa_{ij}$, it substantially complicates the analysis of the measured data with respect to the case of an elongated line-shaped heat source (sensitive to only two cyrstallographic directions simulataneously), as it is evident from Refs. [\onlinecite{Schmidt2008,Feser2012a, Feser2014,Rodin2017,Jiang2017,Jiang2018, Li2018,Rahman2018,Yuan2019,Qian2020, Tang2021,Ly2021}]. 
In addition, for Gaussian or ellipse-shaped small focused spots a precise determination of the spatial intensity distribution of the focused laser spot is required,  which in some situations can be challenging due to the highly asymmetric shape of the heater spot.\cite{Jiang2018, Li2018} 
We note that for elongated line-shaped geometries of the focused spot, the precise shape of the spot must not be necessary known to obtain the thermal properties of the samples (``slope method").
The main objective of the present work is to provide a new technical approach which combines the key advantages of the experimental methods reported in Refs. [\onlinecite{Schmidt2008,Ramu2012,Rodin2017,Feser2012a,Feser2014,Mishra2015a,Jiang2017,Jiang2018, Li2018,Rahman2018,Yuan2019,Qian2020,Tang2021,Ly2021}] but without their drawbacks, i.e., based on a 1-dimensional heat source configuration, which is focused onto the surface of the samples in a contactless fashion.

Here, we developed a novel experimental approach to measure the thermal conductivity tensor in a contactless fashion which is advantageous to study isotropic and anisotropic materials, and thin films. This technique, that we refer to as ``Anisotropic Thermoreflectance Thermometry''(ATT), is based on the geometry used in the 3-omega method, but instead of using an electrical resistor we use a focused laser beam to define a long and narrow line-shaped heat source, hence, taking profit from the many advantages offered by this geometry such as, e.g., the determination of the thermal conductivity through the so-called ``slope method".
We derive the fundamental equations which lead to the solution of the problem for bulk, multilayer systems, and thin films. The excellent experimental performance of this methodology is demonstrated in isotropic bulk materials such as glass, Si, and Ge, as well as for the case of thin films (SiO$_2$/Si).  Finally, we address its applicability to study the thermal conductivity tensor in anisotropic materials such as $\beta$-Ga$_2$O$_3$, and Kapton tape, showing that the present approach requires minimum sample processing and data modeling, but keeping at the same time a rather high experimental accuracy and temperature sensitivity.

\section{Solution for a long and narrow heater line with a Gaussian lateral distribution}

\begin{figure}[t]
\includegraphics[scale=0.7]{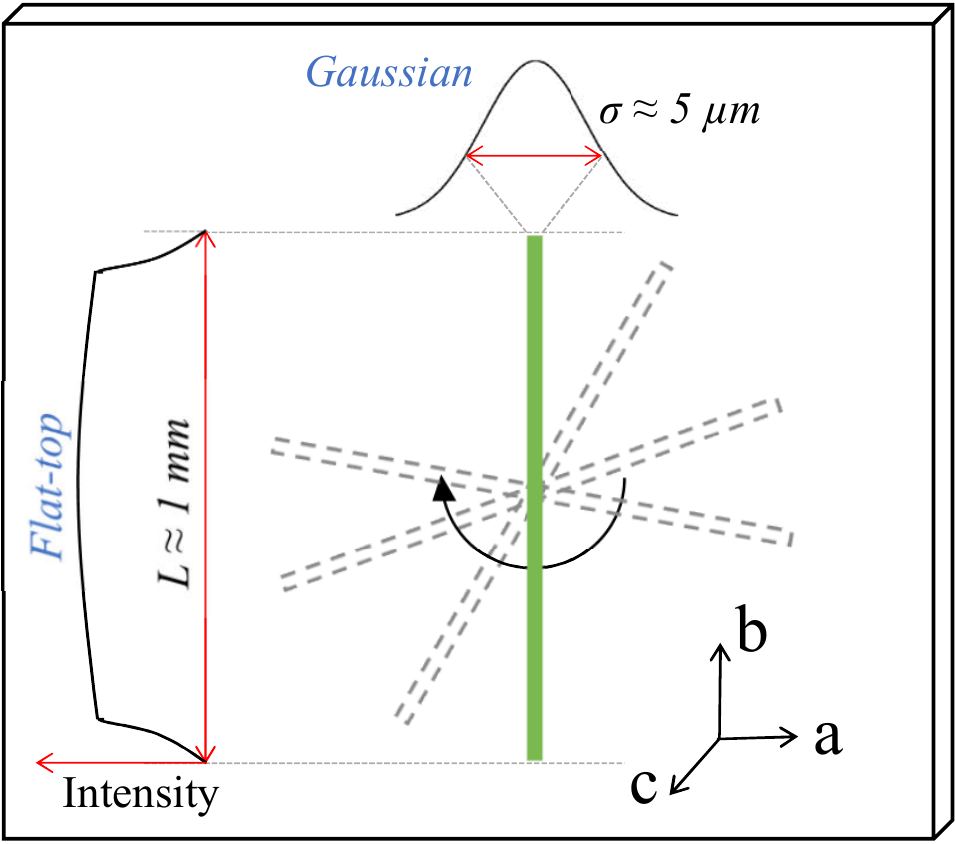}
\caption{Schematic illustration of the studied geometry. The black solid and dashed lines represent the convolution of the pump and probe lasers for different angles with respect to the principal axes of the substrate. The intensity distribution across the region defined by the pump and probe lasers is Gaussian, whereas it has a uniform distribution along the long axis.}
\label{fig1}
\end{figure}

In this section we derive the solution for the temperature response of the system upon a line-shaped thermal excitation. In particular, we provide the mathematical expressions necessary to obtain the thermal conductivity for different geometries of the sample from the frequency-dependent thermal response. The problem we aim to study is the temperature response of a heterostructure subjected to a long and narrow line-shaped heat source. This geometry is similar to that used in the 3-omega method,\cite{Cahill1990, Borca-Tasciuc2001} where a narrow ($\approx$10 $\mu$m) and long ($\approx$1 mm) metallic resistor is deposited onto the surface of the sample, which is used simultaneously as heater and thermometer. However, in our case the heater is optically defined by focusing a Gaussian laser into a line-shaped geometry with uniform power distribution, which is achieved by diffractive beam profile reconstruction. A schematic illustration of the studied geometry is shown in Fig. \ref{fig1}, and the technical details are provided in the experimental section. The intensity distribution of the heat source in the direction across the line is Gaussian, whereas  it is almost uniform along the long axis (flat-top profile). We start by providing the solution for the semi-infinite medium case, which we generalize for the case of an anisotropic multilayer system using the formalism developed by Borca-Tasciuc in Ref. [\onlinecite{Borca-Tasciuc2001}]. We also provide the limit for large thermal penetration depth, which leads to the so-called ``slope method" for a semi-infinite anisotropic sample. 
We recall that this method is particularly attractive since it allows an accurate and rather simple determination of the thermal conductivity since 
$\kappa \propto [\partial T/\partial \ln (\omega )]^{-1}$, hence, a linear fit of the experimental data is sufficient to obtain the thermal conductivity with no additional assumptions and/or numerical modelling. Finally, the solution for the case of a thin film on a semi-infinite substrate is also provided. We note that although the mathematical treatment of this problem is similar to that presented in Refs. [\onlinecite{Cahill1990, Borca-Tasciuc2001}], in our case the heat source intensity distribution in the direction perpendicular to the long axis is Gaussian and not a step function as in the case of the 3-omega method,\cite{Cahill1990} which must be taken into account in order to obtain the correct thermal response of the system.

The solution of the heat equation for the temperature oscillations, $T_\omega(r,t)$, at the surface of a semi-infinite medium under a harmonic excitation produced by a long and narrow surface line heat source\cite{Carslaw1986} is used as starting point:

\begin{equation}
T_\omega(r,t) = \frac{p_0}{\pi\kappa l} K_0(qr)\exp(i 2\omega t) \label{solutioncahill}
\end{equation}
\noindent where $r$ and $t$ are the radial and temporal coordinates, $\omega$ is the angular excitation/detection frequency ($\omega=2\pi f$), $p_0$ is the power dissipated in the heater with length $l$ and it is set to $p_0=1$ W, $\kappa$ is the scalar (isotropic) thermal conductivity of the semi-infinite medium, $\alpha=\kappa/(C_p\rho)$ is the thermal diffusivity, $C_p$ its heat capacity, $\rho$ the density, $K_0$ is the zero-th order modified Bessel function of the second kind, and $q=(i\omega/\alpha)^{1/2}$ is the thermal wavelength with its associated thermal penetration depth, $L_p=1/|q|$, which is defined as the spatial characteristic decay length of $K_0(qr)$. The previous solution is strictly valid for the ideal case of an infinitely long and narrow heat source. The solution for the case of a heat source with finite width is obtained through convolution of Eq. (\ref{solutioncahill})  with the Gaussian power distribution of the heat source. This procedure is further simplified applying the convolution theorem: $\mathcal{F}\{f\cdot g\}=\mathcal{F}\{f\}\cdot \mathcal{F}\{g\}$, where $\mathcal{F}$ represents the Fourier transform, and by taking the inverse Fourier transform of the resulting expression. Thus, we compute the Fourier transform on the spatial coordinate of Eq. (\ref{solutioncahill}) and of the heat source, $P_i(r)=p_iI(\sigma,r)=\frac{p_i}{2\sigma_i \sqrt{2\pi}}\exp[-r^2/(8\sigma_i^2)]$, as:

\begin{eqnarray}
\mathcal{F}_r[T_\omega(r)](\xi) &=& T_\omega(\xi)= \frac{1}{\pi\kappa l}\frac{1}{\sqrt{\xi^2+q^2}} \label{eqn3}\\ 
\mathcal{F}_r[P_i(r)](\xi)&=& P_i(\xi)=p_i\exp[-\sigma_i^2\xi^2/8] \label{eqn4}
\end{eqnarray}

\noindent where the index, $i=p, pr$, accounts for the pump or probe laser, $p_i$ is the absorbed power, and $\sigma_i$ is the $1/e^2$ radius of each laser. The temperature response of the system under a Gaussian heat source is obtained by multiplication of frequency-domain expressions for the ideal line solution (Eq. \ref{eqn3}) and for the heat source power distribution ( Eq. \ref{eqn4}). 

\begin{equation}
 T'_\omega(\xi)=\frac{p_p}{\pi\kappa l_p}\frac{\exp[-\sigma^2_p\xi^2/8]}{\sqrt{\xi^2+q^2}},
\end{equation}

\noindent after taking the inverse Fourier transform we obtain the temperature at a distance $r$ from the center of the heater as:
 
\begin{equation}
  T'_\omega(r)=\frac{p_p}{\pi\kappa l_p}\int_{0}^{\infty}\frac{\exp[-\sigma^2_p\xi^2/8]}{\sqrt{\xi^2+q^2}}\cos(\xi r) d\xi \label{master}
\end{equation}

In order to obtain a mathematical expression which can be directly related to the experiments the thermal response of the system, Eq. (\ref{master}) must be weighted by the normalized intensity distribution of the probe laser, $I_{probe}$, which is also well represented by a Gaussian lineshape. We obtain the following average temperature rise across the heater:

\begin{equation}
  \overline{T'}_\omega=\frac{p_p}{\pi\kappa l_p}\iint_{0}^{\infty}\frac{I_{probe}(r)\exp[-\sigma^2_p\xi^2/8]}{\sqrt{\xi^2+q^2}}\cos(\xi r ) d\xi dr \label{master_convoluted}
\end{equation}
 
\noindent which after taking the integral in the spatial coordinate (i.e., the Fourier transform of the probe laser intensity distribution, see Eq. \ref{eqn4}) leads to the frequency-dependent temperature response of the system:

\begin{equation}
 \overline{T'}_\omega =\frac{p_p}{\pi\kappa l_p}\int_{0}^{\infty}\frac{\exp[-(\sigma_{p}^2+\sigma_{pr}^2) \xi^2/8]}{\sqrt{\xi^2+q^2}} d\xi \label{solution}
\end{equation}

%
%
%

Equation (\ref{solution}) can be numerically solved to render the complex thermal response of the specimen as a function of the excitation frequency, $f$ ($\omega=2\pi f$). Of particular interest, we note that the low frequency limit of Eq. (\ref{solution}) leads to a similar solution as the one obtained in the case of the electrical 3-omega method, i.e., when the thermal penetration depth is much larger than the size of the heater,
$L_p=1/|q|=\sqrt{\alpha/(\omega)} \gg \sigma_{p}$. In this case, the exponential function in Eq. (\ref{solution}) can be expanded to the first non-trivial order, and the upper limit of the integral is set to $1/\sigma_p$ ($L_p\gg \sigma_{p}$). The resultant expression for the temperature response is: 

\begin{equation}
\overline{T'}_\omega =\frac{p_p}{\pi\kappa l_p}\int_{0}^{1/\sigma_p} \frac{1-(\sigma_{p}^2+\sigma_{pr}^2) \xi^2/8 }{\sqrt{\xi^2+q^2}} d\xi \label{solution2}
\end{equation}

\noindent which can be further simplified by setting $\sigma_{pr}=\sigma_{p}$, and by taking the definite integral as:

\begin{equation}
  \overline{T'}_\omega=\frac{p_p}{\pi\kappa l_p}\left[\ln \left( \frac{\sqrt{{\alpha/i}}}{\sigma} \right)-\frac{1}{2}\ln({\omega})-1/8+\ln(2)\right]
\end{equation}

\noindent after rearranging and using the complex relation $\ln(-i)=i\pi/2$, we obtain the following expression:

\begin{equation}
  \overline{T'}_\omega=\frac{p_p}{2\pi\kappa l_p}\left[
  \ln \left( \frac{{{\alpha}}}{\sigma^2} \right)
  -\ln({\omega})-1/4+2\ln(2)-i\frac{\pi}{2}\right] \label{solution3}
\end{equation}

\noindent where the frequency dependence is given solely by the second term on the right-hand side of the previous equation. We remark that the solution obtained  for $L_p\gg \sigma$, is similar to that obtained for the case of the 3-omega method, which is a consequence of the similar line-shaped heater geometry used in both cases.

Finally, Eq. (\ref{solution}) can be generalized to the case of a multilayered system through the model developed by Borca-Tarsciuc.\cite{Borca-Tasciuc2001} We reproduce here only the key ingredients which lead to the final expression for the temperature field. For a detailed derivation we refer the reader to the original publication:\cite{Borca-Tasciuc2001}

\begin{equation}
\overline{T'}_\omega =\frac{p_p}{\pi\kappa^{\perp}_1 l_p}\int_{0}^{\infty}\frac{\exp[-(\sigma_{p}^2+\sigma_{pr}^2) \xi^2/8]}{A_1(\xi)B_1(\xi)} d\xi \label{solution_multilayers}
\end{equation}

\noindent where $A_1$ and $B_1$ are defined as:

\begin{eqnarray}
A_{j-1}=\frac{A_j\frac{\kappa^\perp_j B_j}{\kappa^\perp_{j-1}B_{j-1}}-\tanh(B_{j-1}d_{j-1})}{1-A_j\frac{\kappa^\perp_{j}B_j}{\kappa^\perp_{j-1}B_{j-1}}\tanh(B_{j-1}d_{j-1})},\hspace{5mm}j= (2,..., n)
\end{eqnarray}
\begin{eqnarray}
B_j &=& \sqrt{\frac{\kappa^{\parallel}_j}{\kappa^\perp_j}\xi^2+\frac{2C_j\rho_j(\omega)}{\kappa^\perp_j}}
\end{eqnarray}

and,

\begin{equation}
     A_n = -\tanh(B_n d_n)^s 
\end{equation}

\noindent where $n$ is the number of layers counting from the top surface, i.e., $n=1$ at the surface where the pump and probe lasers are focused, and $n=$ ``total number of layers" at the bottom layer. The parameter $s$ sets the type of boundary condition at the bottom layer with $s=0$ for a semi-infinite substrate. When the substrate thickness is finite, $s=1$, for adiabatic boundary conditions, and $s=-1$, for the case of isothermal boundary conditions. Finally, the thermal boundary resistance between two layers, $K_{TBR}$, is modeled using the usual assumptions, i.e., a  1 nm thick layer with a small heat capacity (e.g, $C_p \approx$ 1 JKg$^{-1}$C$^{-1}$).

The solution for the temperature oscillations for the multilayer case given by Eq. (\ref{solution_multilayers}) includes the in-plane ($\kappa^{\parallel}$, $c$-plane in Fig. \ref{fig1}), and cross-plane ($\kappa^{\perp}$, $c$-axis in Fig. \ref{fig1}) components of the thermal conductivity. Thus, rotating the sample with respect to the heat source long axis leads to the projection of $\kappa_{ij}$ into the $c$-plane. Finally, the low frequency limit (large penetration depth) of Eq. (\ref{solution_multilayers}) for the case of an anisotropic semi-infinite medium leads to an expression similar to Eq. (\ref{solution3}) as shown in Ref. [\onlinecite{Borca-Tasciuc2001}]:

\begin{equation}
  \overline{T'}_\omega=\frac{p_p}{2\pi l_p \sqrt{\kappa^{\perp}\kappa^{\parallel}}}\left[
  \ln \left( \frac{{{\kappa^{\parallel}\alpha}}}{\kappa^{\perp}\sigma^2} \right)
  -\ln({\omega})-1/4+2\ln(2)-i\frac{\pi}{2}\right] \label{solution_slope_anisotropic}
\end{equation}
where $\kappa^{\parallel}$ is an arbitrary direction in the $c$-plane as defined in Fig. \ref{fig1}, which is perpendicular to the direction defined by the heat source. The previous equation implies that for the case of anisotropic substrates, the thermal conductivity extracted from the slope method senses simultaneously the cross-plane and in-plane components,  $\kappa_{eff}=\sqrt{\kappa^{\perp}\kappa^{\parallel}}$. We note that although the complete solution to the multilayer case is given by Eq. (\ref{solution_multilayers}), the low frequency approximation considerably simplifies the data analysis process.

Finally, an expression for the thermal conductivity of a thin film with thickness $d$ deposited on a substrate is provided. We note that in the lower frequency limit (large thermal penetration depth) the thin film can be regarded as a thermal interface resistance and, hence, it does not introduce any frequency-dependent response. This arises from the fact that the thermal penetration depth in the low frequency range, $L_p=|q|^{-1}=(i\omega/\alpha)^{-1/2}$, is typically much larger than the typical thickness of the thin films ($\approx$ 100 nm). The previous argument is strictly valid in the 1-dimensional (1D) heat flow limit, for which the thermal conductivity of the substrate is larger than the thermal conductivity of the thin film, i.e., the substrate behaves as a heat sink. To obtain an expression for the thermal conductivity of the thin film, the area of the Gaussian laser spot is used to define the thermal interface resistance, $\int_{-\infty}^{\infty}\exp{(-2r^2/\sigma)}dr=\sigma\sqrt{\pi/2}$. Thus, the thermal conductivity of the thin film is:

\begin{equation}
    \kappa_{TF}=\frac{Pd}{l_p\sigma{\Delta T}_{TF}\sqrt{\pi/2}}
    \label{equation_TF}
\end{equation}

\noindent where ${\Delta T}_{TF}$ is the frequency independent temperature rise due to the presence of the thin film with respect to the response of the bare substrate (both including the Au transducer).

\begin{figure}[t]
\includegraphics[scale=0.42]{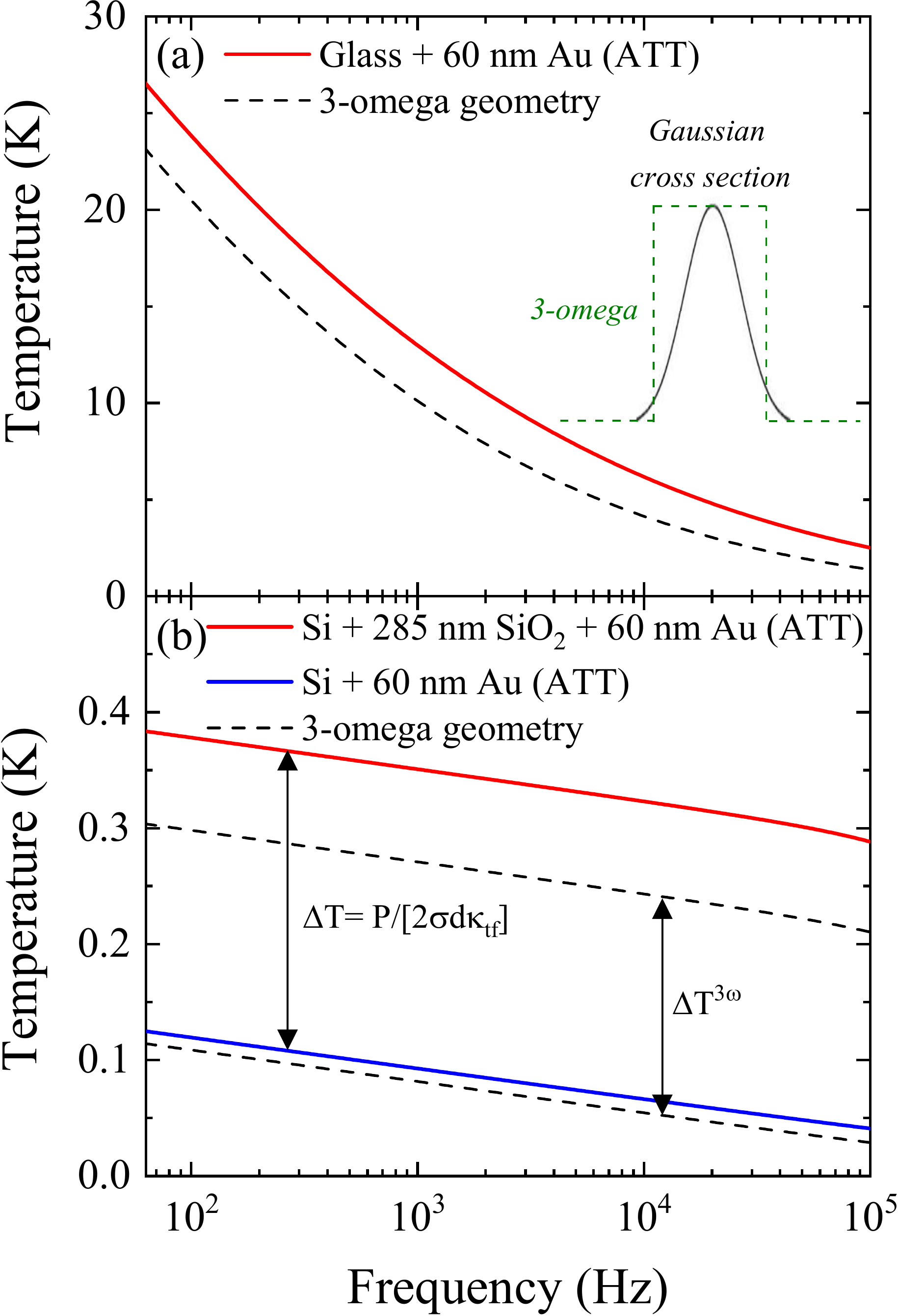}
\caption{Numerical simulations of the maximum temperature oscillations obtained using Eq. (\ref{solution_multilayers}) are shown in full lines. (a) Glass/Au (bulk glass; 60 nm Au). (b) Si/Au (bulk Si; 60 nm Au), and Si/SiO$_2$/Au (bulk Si; 285 nm SiO$_2$; 60 nm Au). For all simulations we used 10 mW of heater power, L = 1 mm for the length of the line-shaped heater, and Gaussian cross sections
$\sigma(r=1/e^2)= $ 5 $\mu$m for the pump and probe lasers. The calculated response for each sample assuming a step-function as the heater profile as used in a 3-omega experiment using the same power density is shown in dashed line. }
\label{fig2}
\end{figure}

We summarize as follows the equations which apply to each case depending on the geometry of the sample, including a short description of the key assumptions:

\begin{enumerate}
    \item {\it Isotropic substrate:}(general solution) $\rightarrow$ Eq. (\ref{solution}) \\
    The solution can be obtained by numerical integration and is valid for the whole frequency range provided that the substrate can be considered as semi-infinite.
    \item {\it Isotropic substrate} ($L_p=|q|^{-1} \gg \sigma$): $\rightarrow$ Eq. (\ref{solution3}) \\
    The solution is valid when the thermal penetration depth ($|q|^{-1}$) is much larger than the width, $\sigma$, of the Gaussian heat source. The thermal conductivity is obtained from the slope method with $\kappa=[P_p/(2\pi l_p)](\partial T/\partial \ln{(\omega)})^{-1}$.                                                                 
    \item {\it Anisotropic substrate} (general solution): $\rightarrow$ Eq. (\ref{solution_multilayers}) \\
    The solution can be obtained by numerical integration and is valid for the whole frequency range provided that the substrate can be considered as semi-infinite.
    \item {\it Anisotropic substrate} ($L_p=|q|^{-1} \gg \sigma$): $\rightarrow$ Eq. (\ref{solution_slope_anisotropic})\\
     The solution is valid when the thermal penetration depth ($|q|^{-1}$) is much bigger than the width, $\sigma$, of the Gaussian heat source. The effective thermal conductivity is obtained from the slope method as $\kappa_{eff}=[P_p/(2\pi l_p)](\partial T/\partial \ln{(f)})^{-1}$, and it is related to the components as  $\kappa_{eff}=\sqrt{\kappa^{\perp}\kappa^{\parallel}}$.
    \item {\it Thin films on substrate}$: \rightarrow$ Eq. (\ref{equation_TF}) \\
    This expression is valid when the thermal conductivity of the substrate is larger than the thermal conductivity of the thin film; rule of thumb $\rightarrow$  $\kappa_{substrate} \gtrapprox 10\kappa_{TF}$.
\end{enumerate}

Figure \ref{fig2} displays the numerical solution of Eq. (\ref{solution}) for the case of glass and Si substrates, as well as the numerical solution of Eq. (\ref{solution_multilayers}) for a SiO$_2$/Si thin film using 10 mW of heater power. We plot the frequency-dependent maximum temperature rise obtained using the present method (ATT), as well as the corresponding solution  for the 3-omega geometry, which would correspond to an illumination beam with square section in the sort axis (see inset of Fig. \ref{fig2}a). 
Figure \ref{fig2}a displays the case of a glass substrate coated with 60 nm of a Au transducer. As expected, the temperature response approaches a linear dependence (in logarithmic frequency scale) for rather low frequencies (large thermal penetration depth). Within the lower frequency range, the thermal conductivity can be directly obtained through the ``slope method", i.e., $\kappa \propto [\partial T/\partial \ln (\omega )]^{-1}$.  For higher frequencies, the thermal penetration depth decreases and becomes comparable with the width of the heat source, hence, leading to deviations from the linear behaviour (in logarithmic frequency scale) observed at lower frequencies. Figure \ref{fig2}b displays the amplitude of the temperature oscillations for a Si substrate as well as for a 285 nm thick SiO$_2$ thin film deposited on a Si substrate. For Si, the maximum temperature depends linearly on $\ln(\omega)$, which is a direct consequence of the large thermal penetration depth originated from its large thermal conductivity (or equivalently, thermal diffusivity $\rightarrow \alpha=\kappa C_p^{-1}\rho^{-1}$). It is interesting to note that in all studied cases the temperature response is similar to that obtained from the 3-omega method (dashed lines in Fig. \ref{fig2}), however, for similar excitation spot sizes the magnitude of the temperature response of the present approach is always larger as compared to the case of the 3-omega method. This originates from the fact that the power density is larger by a factor, $\sqrt{8/\pi}$, for a Gaussian intensity distribution with radius, $\sigma$,  as compared to a step-function distribution with half with, $b$, when $\sigma \approx b$.

\section{Experimental Approach}
\begin{figure}[t]
\includegraphics[scale=1]{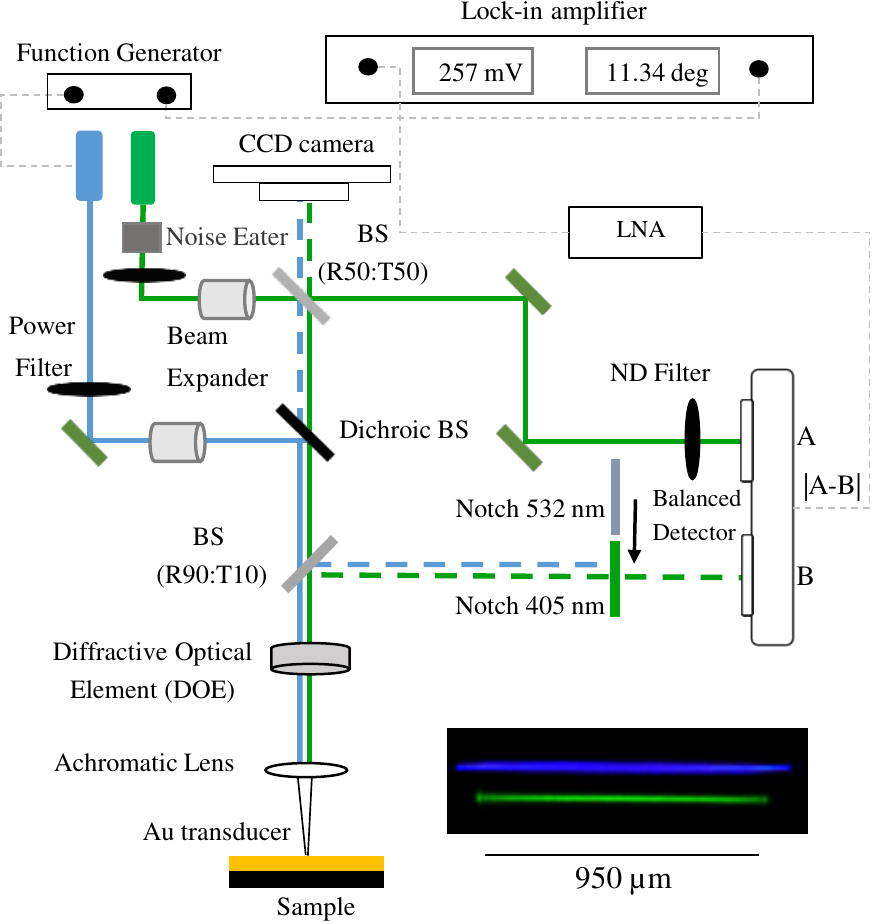}
\caption{Schematics of the experimental arrangement used in the present experiments indicating each of the optical components. An optical image of of the pump (blue) and probe (green) lasers after intensity reconstruction is also shown.}
\label{fig3}
\end{figure}

The experimental setup used in the present experiments is shown in the schematic illustration of Fig. \ref{fig3}. The experiments are based on a frequency-domain pump-and-probe concept where the wavelength of the pump laser was set to $\lambda^{pump}=405$ nm (Omicron A350) and the probe laser wavelength was $\lambda^{probe}=532$ nm (Cobolt SAMBA series). Whereas the pump laser power was modulated with an external harmonic wave generator (Rigol DG5332) between 63 Hz and 100 kHz, the incident probe laser intensity was constant in time. The power of the probe laser was kept low, in order avoid any heating effects, typically of the order of hundreds of $\mu$W. On the other hand the pump laser power was increased until a temperature rise of the order of several degrees was observed, depending on the thermal conductivity of the sample. The typical power used for the heater laser was in all cases $\approx$ 50 mW, however, this quantity was precisely measured for each sample and optical alignment conditions. Both lasers were coupled into the main optical axis using several beam splitters (BS), and dichroic beam splitters. The diameter of the pump and probe beams at the laser output was 1100 $\mu$m and 800 $\mu$m, respectively, and their intensity distribution was Gaussian. Two beam expanders were placed in the optical path of the lasers since the beam diameter must be adjusted precisely for successful diffractive beam reconstruction, which was achieved through a custom designed holographic diffractive optical element (DOE) purchased from HOLO/OR Ltd. This element was placed just before an achromatic focusing lens (f = 50 mm). The DOE is one of the key elements of the experimental setup, since it is responsible for homogenizing the intensity of the pump and probe lasers, i.e. the Gaussian distribution is converted through diffraction into a line-shaped spot with uniform intensity distribution along the long axis, and with Gaussian distribution in the perpendicular direction. Figure \ref{fig3} also displays an optical image of the pump and probe lasers after intensity reconstruction using the DOE. We note that although there are several ways to convert a Gaussian beam into a flat-top distribution such as, e.g., using a fiber with square core to induce mode mixing, or simply using an optical diffuser in combination with a cylindrical lens, we have observed that using a DOE leads to the best results in terms of spot shape and size, and output pump laser power. The back reflection of the lasers after focusing on the sample were coupled into the detection arm using a BS, and the laser wavelength was selected by two mechanically controlled notch-filters. As detector we used a large area (5x5 mm$^2$) AC-coupled balanced detector (Thorlabs PDB210A/M-AC-SP). A sample of the probe laser (532 nm) was independently coupled to the second detector input. The output of the AC-coupled balanced detector is sensitive to the difference between both inputs. The purpose of such detector is reducing the intrinsic noise within the probe laser, typically 0.2$\%$. When the sample is excited by the harmonic pump laser, the output of the detector was a frequency modulated signal arising from the thermal oscillations on the surface of the sample. We note that using such a balanced detector scheme is convenient mostly in the lower frequency range, $f <5$ kHz, since the total noise is dominated by $f^{-1}$ noise. In order to minimize the laser noise of the optical path of the signal and reference lasers were kept similar within 10 cm, and the input power on both detectors was balanced. The pump laser was also coupled to the detection system in order to obtain a reference phase for the thermal signal, i.e., first the pump laser signal (amplitude and phase) was measured by the lock-in amplifier after which the mechanical notch-filter were switched, and the amplitude and phase of the thermal signal was obtained. This process was sequentially repeated for each excitation frequency. 

In order to convert the voltage measurement (typically in V units) to temperature, several factors must be taken into account. First, the amplitude of the reflectivity oscillations measured with the lock-in amplifier ($V_{AC}$) must be normalized by the DC reflectivity at the probe wavelength (532 nm). This process must be carefully done since the balanced detector and the low noise amplifier involve several amplification stages of the modulated signal (thermal signal). The resulting voltage ratio $V_{AC}/V_{DC}=\Delta R/R_0$. The reflectivity change, $\Delta R/R_0$, is converted to temperature by using the temperature coefficient of reflectivity of the Au transducer. 

Regarding the phase lag of the optical reflectivity with respect to the thermal harmonic excitation, it can be shown by solving Eq. (\ref{solution}) that for substrates with rather low thermal conductivity such as glass, the phase lag is almost independent on the thermal boundary resistance (TBR) between the Au transducer and the substrate. However, for substrates with higher thermal conductivity such as Si, the phase lag is affected by the TBR and, hence, this method can be also used to determine the TBR. Note that in both previous cases the geometry of the heat source also affects the phase lag and, thus, its dimensions must be precisely measured.
With respect to the absolute value of the temperature rise, it can be also shown through the solution of Eq. (\ref{solution}), that its logarithmic frequency derivative remains approximately unaffected, as expected for the ``slope method".


The photothermal response of the Au transducer was calibrated studying a glass control sample whose thermal conductivity was independently measured using the 3-omega method, obtaining a thermal conductivity of 0.96 Wm$^{-1}$K$^{-1}$. This value was used as input calibration to obtain the temperature coefficient of reflectivity of the Au transducer of the same glass sample through application of the slope method using Eq. (\ref{solution3}). We have obtained, $(1/R_0)[\partial R/ \partial T]=3\times10^{-4}$K$^{-1}$, in good agreement with most reported values for Au.\cite{Favaloro2015} So far, we have not observed an appreciable variation of this coefficient between samples fabricated in different Au evaporation cycles, since we always use the same evaporation conditions.
The estimated temperature resolution of this experimental approach was observed to be as low as $\approx$3 mK, whereas the error in the determination of the thermal conductivity is $\approx$5$\%$. A detailed study of the various error sources will be considered for future study.

\section{Results and Discussion}

We have studied five bulk samples of glass, Si(100), Ge(100), Ga$_2$O$_3(\overline{2}01)$, and Kapton tape, as well as a 285 nm thick SiO$_2$ thin film deposited on a Si(100) substrate. All samples were coated with a 60 nm thick Au thin film (transducer) which was thermally evaporated at a base pressure of 10$^{-6}$ mbar. All measurements were done in ambient conditions. Of particular interest, we investigated a $(\overline{2}01)$ oriented Ga$_2$O$_3$ substrate, which exhibits strong in-plane thermal anisotropy. We show in this section that the present method provides a straightforward means to determine the in-plane components of the thermal conductivity tensor.

\subsection{Isotropic Materials and thin films}
\begin{figure}[t]
\includegraphics[scale=0.45]{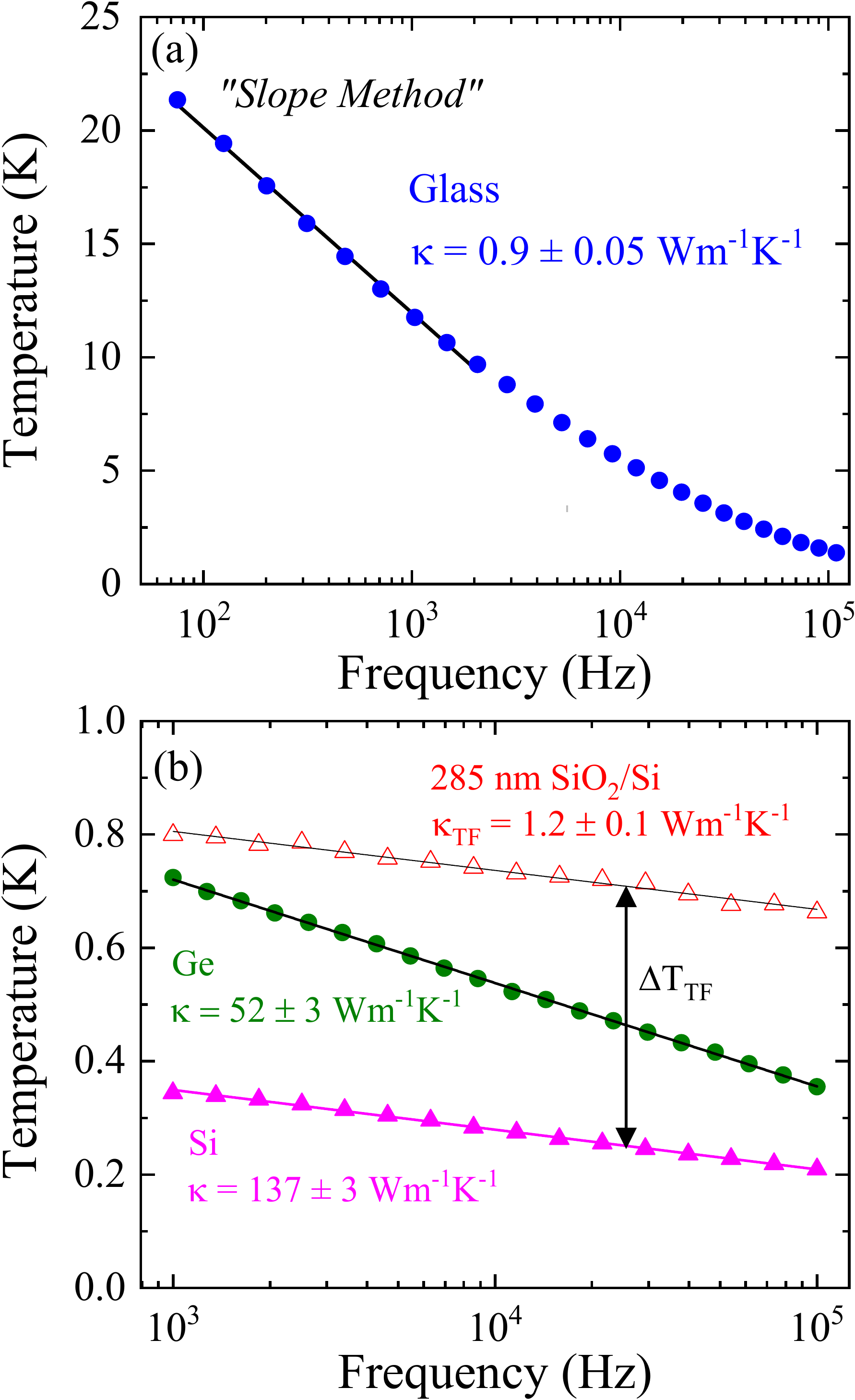}
\caption{Maximum temperature rise as a function of the modulation frequency of the heater for (a) a glass substrate and, (b) Si and Ge substrates, as well as a 285 nm thick SiO$_2$/Si thin film.}

\label{fig4}
\end{figure}

 Figure \ref{fig4} displays the experimental measurements of the amplitude of the temperature oscillations as a function of the excitation frequency for the case of glass, Si, Ge, as well as for 285 nm thick SiO$_2$/Si thin film. For the glass substrate, shown in Fig. \ref{fig4}a, the temperature decays approximately linearly with $\ln(\omega)$. However, as the frequency increases, the frequency-dependent temperature rise gradually deviates from the linear behavior observed at lower frequencies, in good agreement with the calculations shown in Fig. \ref{fig2}.  We recall, that the linear dependence observed at lower frequencies allows the determination of the thermal conductivity directly through the ``slope method", i.e., avoiding complex data modeling and numerical fitting of the experimental data. This is a key  aspect since in order to quantitatively model the experimental data, it is necessary to know the precise width of the line-shaped optical heater. However, the slope of the linear dependence observed at lower frequencies is independent to some extent on the width of the heater. A linear fit to the data points in the frequency range up to $\approx$1 kHz, results in a thermal conductivity of 0.96 $\pm$ 0.05 Wm$^{-1}$K$^{-1}$. 

In Figure \ref{fig4}b we show the frequency-dependent temperature response for the case of Si, Ge, and a 285 nm thick SiO$_2$ thin film deposited on a Si(100) substrate. Note that the used frequency range is different as compared to the case of glass (see Fig. \ref{fig4}a), since for the lower frequency range ($f \lessapprox 1$ kHz) the thermal penetration depth for Si and Ge becomes comparable with the thickness of the substrates, hence, the thermal response is influenced by the sample holder.
The amplitude of the temperature rise is observed to be dependent linearly on $\ln(\omega)$ for Si and Ge, which resembles the observations on the glass sample at lower frequencies. However, for Si and Ge the thermal penetration depth, $L_p=1/|q|=\sqrt{\alpha/(\omega)}$, is substantially larger as compared to the case of glass since $L_p \propto \sqrt{\alpha}$, where we recall that, $\alpha=\kappa/(C_p\rho)$, is the thermal diffusivity of the sample. The thermal conductivity of the samples was obtained using the slope method as, 137 $\pm$ 7 Wm$^{-1}$K$^{-1}$ and 52 $\pm$ 3 Wm$^{-1}$K$^{-1}$ for Si and Ge, respectively. The frequency dependence of the temperature rise for the 285 nm thick SiO$_2$/Si thin film is similar to the case of pure Si as shown in Fig. \ref{fig4}b, although a temperature offset arising from the presence of the SiO$_2$ thin film is observed. We recall that, similarly as for the case of the 3-omega method, no frequency dependence is introduced by the presence of the thin film in the temperature rise. In other words, if the thermal penetration depth is much larger than the thickness of the thin films, no frequency dependence of the temperature rise is expected.

\subsection{Anisotropic Materials}

\begin{figure}[t]
\includegraphics[scale=0.45]{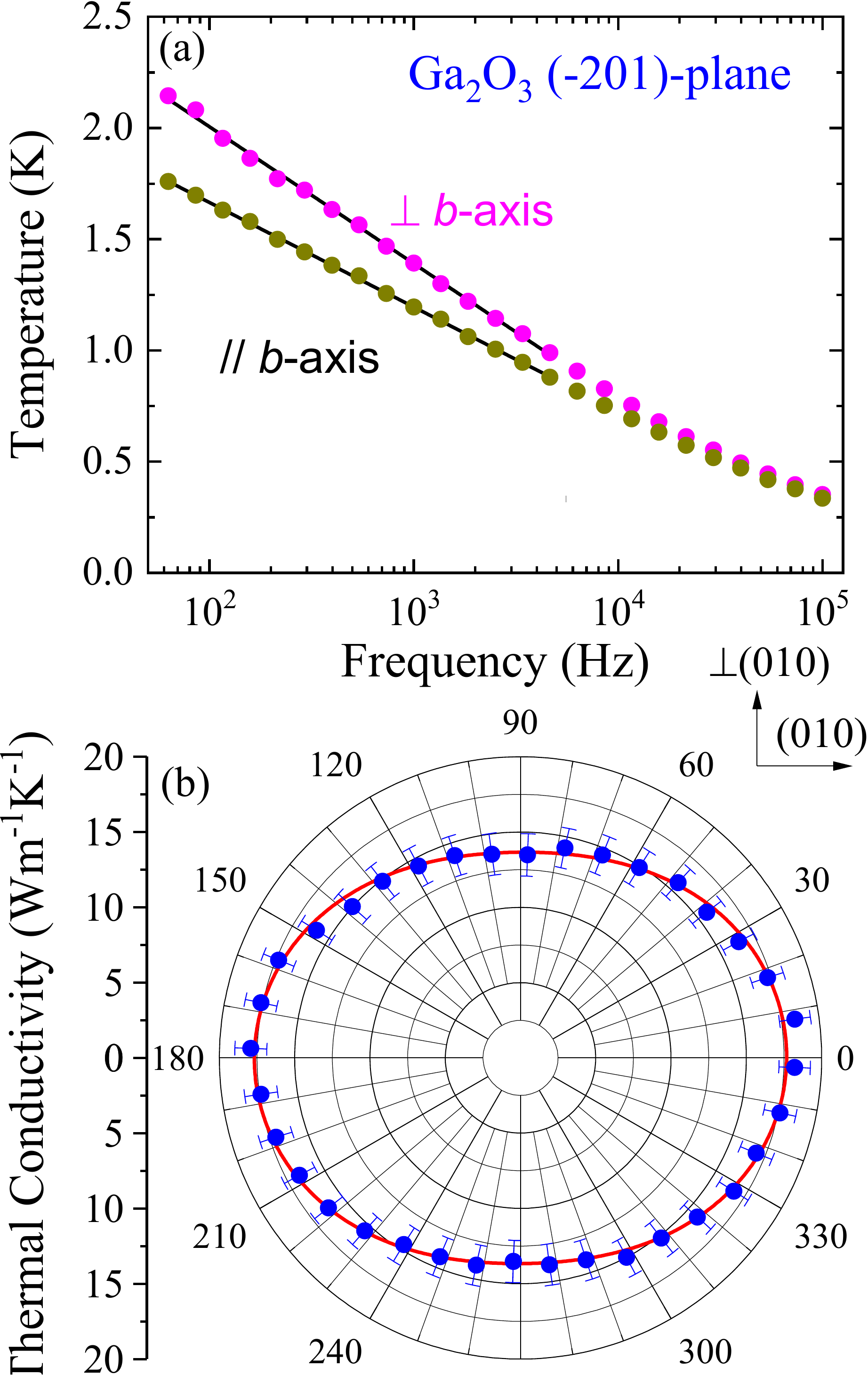}
\caption{(a) Temperature response curves in for $\beta$-Ga$_2$O$_3$ along the directions $b$ and $b^{\perp}$. The frequency range below $\approx$10 kHz exhibits a linear response in logarithmic frequency scale. (b) Polar plot of the thermal conductivity tensor projected onto the $(\overline{2}01)$-plane of a  $\beta$-Ga$_2$O$_3$ substrate. The vertical scale represents the radial values of the thermal conductivity. }
\label{fig5}
\end{figure}

Finally, we demonstrate the strengths of the present approach for the study of the thermal properties of anisotropic materials. As a model system, we chose the sesquioxide $\beta$-Ga$_2$O$_3$, which is the most stable of several Ga$_2$O$_3$ polymorphs and crystallizes in the monoclinic structure (space group C2/m) with pronounced optical, electrical, and thermal anisotropy. Here, we present measurements of a single crystal $\beta$-Ga$_2$O$_3$ substrate with $(\overline{2}01)$-plane. Figure \ref{fig5}a displays the temperature rise as a function of the excitation frequency for two perpendicular in-plane directions corresponding to the $[010]$ and ${\perp}[010]$ crystallographic directions. A different temperature response is observed for these two directions, with a lower temperature rise along the [010] as compared to ${\perp}[010]$. The temperature response is linear in logarithmic scale up to $\approx$5 kHz, whereas for higher frequencies a similar deviation as previously discussed for the case of glass is observed. Thus, the anisotropic thermal conductivity is once more obtained directly by the ``slope method'' in this frequency range, i.e., without the need for numerical fitting.   

\begin{figure}[t]
\includegraphics[scale=0.45]{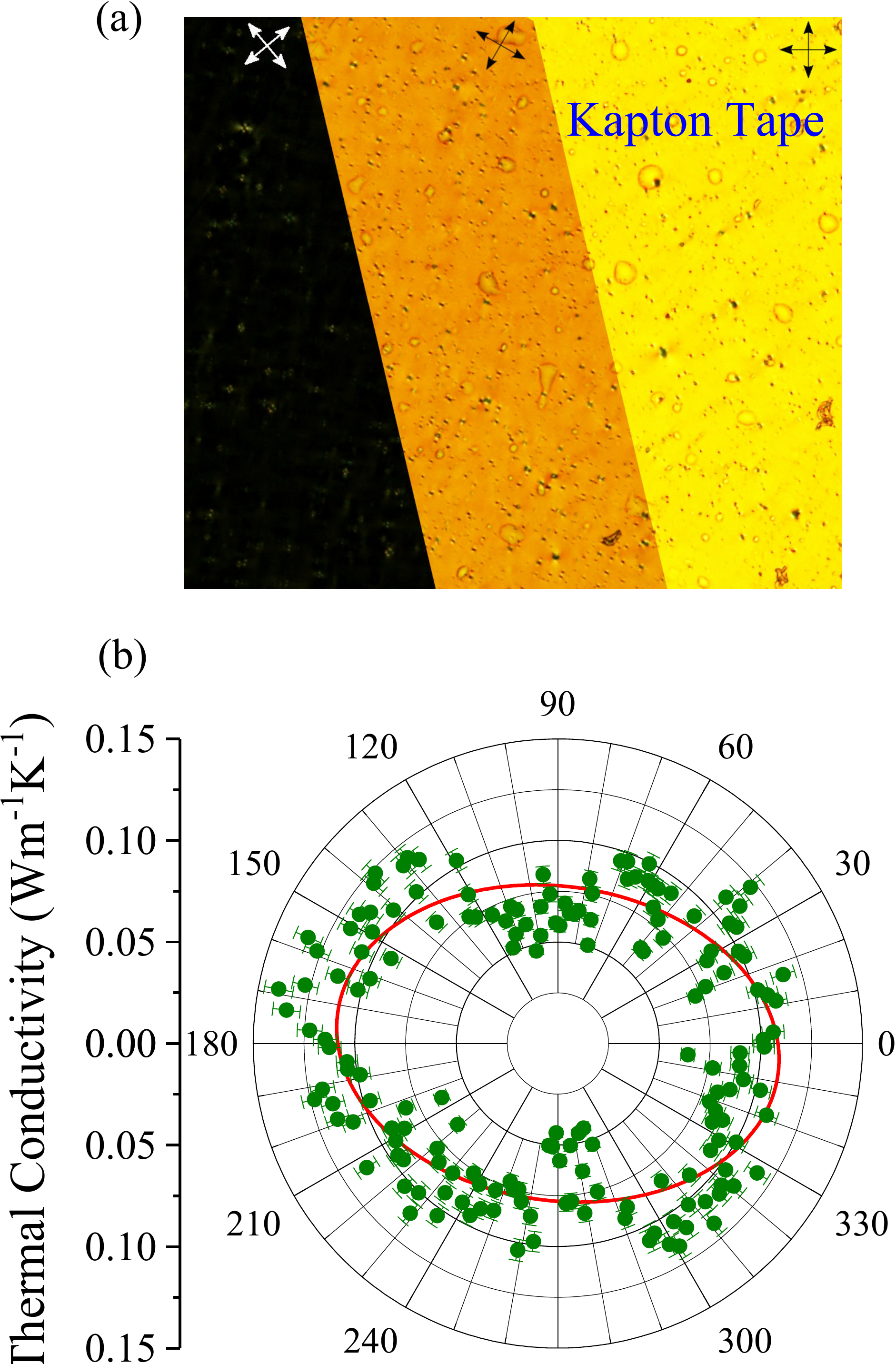}
\caption{(a) Polarized optical image of a Kapton tape showing its optical anisotropy. (b) Polar plot of the thermal conductivity tensor projected onto the plane of the sample. The vertical scale represents the radial values of the thermal conductivity, and the full line is least-squares fits using an elliptical function for the in-plane thermal conductivity distribution.}
\label{fig6}
\end{figure}

The complete determination of the thermal anisotropy within the $(\overline{2}01)$-plane is then achieved by rotating the sample with respect to the optical line-shaped heater. Figure \ref{fig5}b displays the in-plane thermal conductivity as a function of the rotation angle using 10$^{\circ}$ steps. We recall that for anisotropic samples, the temperature response at each orientation of the line-shaped heater with respect to the principal axes is determined by the values of the thermal conductivity in both directions perpendicular to the line heater, simultaneously. The direction parallel to the line heater has a uniform temperature distribution and, thus, no heat flow is present in this direction. The thermal conductivity along the direction $j$, is thus obtained using the following expression (see section II):

\begin{equation} 
\kappa^j_{eff}=\sqrt{\kappa^{[\overline{2}01]}\kappa^{j}}
\label{aniso}
\end{equation}

\noindent where $\kappa^{[\overline{2}01]}$ is the thermal conductivity along the $[\overline{2}01$] direction, and $\kappa^j_{eff}$ is the measured effective thermal conductivity along the direction $j$. The in-plane thermal anisotropy factor, i.e., the ratio between the maximum and minimum $\kappa$ within the $(\overline{2}01)$-plane, can be directly obtained from Fig. \ref{fig5} (see eq. \ref{aniso}) as $\eta=1.36$.
However, in order to quantitatively obtain the projection of the thermal conductivity tensor on the $(\overline{2}01)$-plane, the out-of-plane value of the thermal conductivity $\kappa^{[\overline{2}01]}$ must be known. 
In order to obtain this value, we have studied the same sample using conventional frequency-domain thermoreflectance based on a Gaussian spot (with spot size of $\approx$ 16 $\mu$m). We have focused on the higher frequency range ($f>$ 100 kHz), where the thermal response is 1-dimensional, hence, domintated by $\kappa^{[\overline{2}01]}$ for the studied sample. We obtained $\kappa^{[\overline{2}01]}=$ 9.1 Wm$^{-1}$K$^{-1}$ in good agreement with the value reported in Ref. [\onlinecite{Jiang2018}]. We note that in order to obtain  all components of the thermal conductivity tensor, additional measurements of different crystallographic planes are required. In fact, the determination of the full thermal conductivity tensor for $\beta-$Ga$_2$O$_3$ using the here presented anisotropic thermoreflectance thermometry technique will be published in a forthcoming work.

Finally, in Figure \ref{fig6} we show the optical and thermal response of a piece of Kapton tape. We aim to demonstrate that even if the quality of the surface is far from ideal (arising in this case from the fabrication process), it is still possible to determine in-plane thermal anisotropy. In particular, the fabrication process of Kapton tapes involves stretching of the material in one direction. 
Figure \ref{fig6}a
displays a polarized optical image of the sample, where surface inhomogeneities, roughness and bubbles can be appreciated. Moreover, the principal optical axis can be easily identified by optical polarization. Accordingly, the thermal anisotropy scan exhibits anisotropy in a direction coincident with the optical axis of the material. 
We note that the dispersion of the points within Figure \ref{fig6}b does not only arise from experimental errors (indicated in Fig. \ref{fig6}), but also from defects inherently present at the surface of the sample, which can affect the temperature distribution.

\section{Conclusions}

We have developed a novel contactless experimental approach suitable to study anisotropic thermal transport based on thermoreflectance thermometry. We demonstrated its applicability to the case of isotropic (Si, Ge) and anisotropic ($\beta$-Ga$_2$O$_3$) and Kapton bulk samples, as well as to thin films (SiO$_2$/Si). The  experimental accuracy of the method in the determination of the thermal conductivity is about $\approx$ 5\%, whereas the temperature resolution is 
$\Delta$T $\approx$ 3mK. 
The method combines all the advantages offered by non-contact approaches, but keeping  mathematical simplicity in the data evaluation process. In fact, a simple linear fit to the frequency dependent temperature response (``slope method") is sufficient to obtain the thermal conductivity of the samples, similar as in the case of the 3-omega method. 
The key advantage of this approach from the thermal perspective lies in the geometry of the heat source, which provides simultaneous sensitivity to two crystallographic directions (perpendicular to the heater line), whereas it is insensitive to the direction parallel to the heater. This ability, together with its contactless implementation, allows to probe all in-plane directions by continuous rotation of the sample. Hence, the method is particularly convenient for measuring the projection of the thermal conductivity tensor onto the plane corresponding to the surface of the samples. We envisage that the present experimental approach provides a new alternative to accurately determine the thermal conductivity tensor in anisotropic materials.

\section{Acknowledgements}
We acknowledge financial support from the Spanish Ministerio de Econom\'ia, Industria y Competitividad for its support through grant CEX2019-000917-S (FUNFUTURE) in the framework of the Spanish Severo Ochoa Centre of Excellence program, and grants PID2020-119777GB-I00 (THERM2MAIN), PGC2018-095411-B-100 (RAINBOW), MAT2017-90024-P (TANGENTS)-EI/FEDER, and 2020AEP141; the Generalitat de Catalunya through grants 2017SGR488 and AGAUR 2018 PROD 00191; and from the European Research Council (ERC) under grant agreement no. 648901. 
{\bf Data and materials availability:} All data needed to evaluate the conclusions in the paper are present in the paper and/or the Supplementary Materials. Additional data related to this paper may be requested from the authors.

\bibliography{cylindrical_fdtr_RSI_v3}

\begin{thebibliography}{33}%
\makeatletter
\providecommand \@ifxundefined [1]{%
 \@ifx{#1\undefined}
}%
\providecommand \@ifnum [1]{%
 \ifnum #1\expandafter \@firstoftwo
 \else \expandafter \@secondoftwo
 \fi
}%
\providecommand \@ifx [1]{%
 \ifx #1\expandafter \@firstoftwo
 \else \expandafter \@secondoftwo
 \fi
}%
\providecommand \natexlab [1]{#1}%
\providecommand \enquote  [1]{``#1''}%
\providecommand \bibnamefont  [1]{#1}%
\providecommand \bibfnamefont [1]{#1}%
\providecommand \citenamefont [1]{#1}%
\providecommand \href@noop [0]{\@secondoftwo}%
\providecommand \href [0]{\begingroup \@sanitize@url \@href}%
\providecommand \@href[1]{\@@startlink{#1}\@@href}%
\providecommand \@@href[1]{\endgroup#1\@@endlink}%
\providecommand \@sanitize@url [0]{\catcode `\\12\catcode `\$12\catcode
  `\&12\catcode `\#12\catcode `\^12\catcode `\_12\catcode `\%12\relax}%
\providecommand \@@startlink[1]{}%
\providecommand \@@endlink[0]{}%
\providecommand \url  [0]{\begingroup\@sanitize@url \@url }%
\providecommand \@url [1]{\endgroup\@href {#1}{\urlprefix }}%
\providecommand \urlprefix  [0]{URL }%
\providecommand \Eprint [0]{\href }%
\providecommand \doibase [0]{http://dx.doi.org/}%
\providecommand \selectlanguage [0]{\@gobble}%
\providecommand \bibinfo  [0]{\@secondoftwo}%
\providecommand \bibfield  [0]{\@secondoftwo}%
\providecommand \translation [1]{[#1]}%
\providecommand \BibitemOpen [0]{}%
\providecommand \bibitemStop [0]{}%
\providecommand \bibitemNoStop [0]{.\EOS\space}%
\providecommand \EOS [0]{\spacefactor3000\relax}%
\providecommand \BibitemShut  [1]{\csname bibitem#1\endcsname}%
\let\auto@bib@innerbib\@empty
\bibitem [{\citenamefont {Renteria}\ \emph {et~al.}(2015)\citenamefont
  {Renteria}, \citenamefont {Ramirez}, \citenamefont {Malekpour}, \citenamefont
  {Alonso}, \citenamefont {Centeno}, \citenamefont {Zurutuza}, \citenamefont
  {Cocemasov}, \citenamefont {Nika},\ and\ \citenamefont
  {Balandin}}]{Renteria2015}%
  \BibitemOpen
  \bibfield  {author} {\bibinfo {author} {\bibfnamefont {J.~D.}\ \bibnamefont
  {Renteria}}, \bibinfo {author} {\bibfnamefont {S.}~\bibnamefont {Ramirez}},
  \bibinfo {author} {\bibfnamefont {H.}~\bibnamefont {Malekpour}}, \bibinfo
  {author} {\bibfnamefont {B.}~\bibnamefont {Alonso}}, \bibinfo {author}
  {\bibfnamefont {A.}~\bibnamefont {Centeno}}, \bibinfo {author} {\bibfnamefont
  {A.}~\bibnamefont {Zurutuza}}, \bibinfo {author} {\bibfnamefont {A.~I.}\
  \bibnamefont {Cocemasov}}, \bibinfo {author} {\bibfnamefont {D.~L.}\
  \bibnamefont {Nika}}, \ and\ \bibinfo {author} {\bibfnamefont {A.~A.}\
  \bibnamefont {Balandin}},\ }\href@noop {} {\bibfield  {journal} {\bibinfo
  {journal} {Advanced Functional Materials}\ }\textbf {\bibinfo {volume}
  {25}},\ \bibinfo {pages} {4664} (\bibinfo {year} {2015})}\BibitemShut
  {NoStop}%
\bibitem [{\citenamefont {Jang}\ \emph {et~al.}(2015)\citenamefont {Jang},
  \citenamefont {Wood}, \citenamefont {Ryder}, \citenamefont {Hersam},\ and\
  \citenamefont {Cahill}}]{Jang2015a}%
  \BibitemOpen
  \bibfield  {author} {\bibinfo {author} {\bibfnamefont {H.}~\bibnamefont
  {Jang}}, \bibinfo {author} {\bibfnamefont {J.~D.}\ \bibnamefont {Wood}},
  \bibinfo {author} {\bibfnamefont {C.~R.}\ \bibnamefont {Ryder}}, \bibinfo
  {author} {\bibfnamefont {M.~C.}\ \bibnamefont {Hersam}}, \ and\ \bibinfo
  {author} {\bibfnamefont {D.~G.}\ \bibnamefont {Cahill}},\ }\href@noop {}
  {\bibfield  {journal} {\bibinfo  {journal} {Advanced Materials}\ }\textbf
  {\bibinfo {volume} {27}},\ \bibinfo {pages} {8017} (\bibinfo {year}
  {2015})}\BibitemShut {NoStop}%
\bibitem [{\citenamefont {Lee}\ \emph {et~al.}(2015)\citenamefont {Lee},
  \citenamefont {Yang}, \citenamefont {Suh}, \citenamefont {Yang},
  \citenamefont {Lee}, \citenamefont {Li}, \citenamefont {Choe}, \citenamefont
  {Suslu}, \citenamefont {Chen}, \citenamefont {Ko}, \citenamefont {Park},
  \citenamefont {Liu}, \citenamefont {Li}, \citenamefont {Hippalgaonkar},
  \citenamefont {Urban}, \citenamefont {Tongay},\ and\ \citenamefont
  {Wu}}]{Lee2015b}%
  \BibitemOpen
  \bibfield  {author} {\bibinfo {author} {\bibfnamefont {S.}~\bibnamefont
  {Lee}}, \bibinfo {author} {\bibfnamefont {F.}~\bibnamefont {Yang}}, \bibinfo
  {author} {\bibfnamefont {J.}~\bibnamefont {Suh}}, \bibinfo {author}
  {\bibfnamefont {S.}~\bibnamefont {Yang}}, \bibinfo {author} {\bibfnamefont
  {Y.}~\bibnamefont {Lee}}, \bibinfo {author} {\bibfnamefont {G.}~\bibnamefont
  {Li}}, \bibinfo {author} {\bibfnamefont {H.~S.}\ \bibnamefont {Choe}},
  \bibinfo {author} {\bibfnamefont {A.}~\bibnamefont {Suslu}}, \bibinfo
  {author} {\bibfnamefont {Y.}~\bibnamefont {Chen}}, \bibinfo {author}
  {\bibfnamefont {C.}~\bibnamefont {Ko}}, \bibinfo {author} {\bibfnamefont
  {J.}~\bibnamefont {Park}}, \bibinfo {author} {\bibfnamefont {K.}~\bibnamefont
  {Liu}}, \bibinfo {author} {\bibfnamefont {J.}~\bibnamefont {Li}}, \bibinfo
  {author} {\bibfnamefont {K.}~\bibnamefont {Hippalgaonkar}}, \bibinfo {author}
  {\bibfnamefont {J.~J.}\ \bibnamefont {Urban}}, \bibinfo {author}
  {\bibfnamefont {S.}~\bibnamefont {Tongay}}, \ and\ \bibinfo {author}
  {\bibfnamefont {J.}~\bibnamefont {Wu}},\ }\href@noop {} {\bibfield  {journal}
  {\bibinfo  {journal} {Nature Communications}\ }\textbf {\bibinfo {volume}
  {6}} (\bibinfo {year} {2015})}\BibitemShut {NoStop}%
\bibitem [{\citenamefont {Li}\ \emph {et~al.}(2017)\citenamefont {Li},
  \citenamefont {Liu}, \citenamefont {Zheng}, \citenamefont {Yan},
  \citenamefont {Yang}, \citenamefont {Lv}, \citenamefont {Xu}, \citenamefont
  {Wang}, \citenamefont {Lu}, \citenamefont {Chen},\ and\ \citenamefont
  {Zhu}}]{Li2017}%
  \BibitemOpen
  \bibfield  {author} {\bibinfo {author} {\bibfnamefont {X.}~\bibnamefont
  {Li}}, \bibinfo {author} {\bibfnamefont {Y.}~\bibnamefont {Liu}}, \bibinfo
  {author} {\bibfnamefont {Q.}~\bibnamefont {Zheng}}, \bibinfo {author}
  {\bibfnamefont {X.}~\bibnamefont {Yan}}, \bibinfo {author} {\bibfnamefont
  {X.}~\bibnamefont {Yang}}, \bibinfo {author} {\bibfnamefont {G.}~\bibnamefont
  {Lv}}, \bibinfo {author} {\bibfnamefont {N.}~\bibnamefont {Xu}}, \bibinfo
  {author} {\bibfnamefont {Y.}~\bibnamefont {Wang}}, \bibinfo {author}
  {\bibfnamefont {M.}~\bibnamefont {Lu}}, \bibinfo {author} {\bibfnamefont
  {K.}~\bibnamefont {Chen}}, \ and\ \bibinfo {author} {\bibfnamefont
  {J.}~\bibnamefont {Zhu}},\ }\href@noop {} {\bibfield  {journal} {\bibinfo
  {journal} {Applied Physics Letters}\ }\textbf {\bibinfo {volume} {111}},\
  \bibinfo {pages} {163102} (\bibinfo {year} {2017})}\BibitemShut {NoStop}%
\bibitem [{\citenamefont {Romano}\ and\ \citenamefont
  {Kolpak}(2017)}]{Romano2017}%
  \BibitemOpen
  \bibfield  {author} {\bibinfo {author} {\bibfnamefont {G.}~\bibnamefont
  {Romano}}\ and\ \bibinfo {author} {\bibfnamefont {A.~M.}\ \bibnamefont
  {Kolpak}},\ }\href@noop {} {\bibfield  {journal} {\bibinfo  {journal}
  {Applied Physics Letters}\ }\textbf {\bibinfo {volume} {110}},\ \bibinfo
  {pages} {093104} (\bibinfo {year} {2017})}\BibitemShut {NoStop}%
\bibitem [{\citenamefont {Jiang}\ \emph
  {et~al.}(2018{\natexlab{a}})\citenamefont {Jiang}, \citenamefont {Qian},
  \citenamefont {Yang},\ and\ \citenamefont {Lindsay}}]{Jiang2018b}%
  \BibitemOpen
  \bibfield  {author} {\bibinfo {author} {\bibfnamefont {P.}~\bibnamefont
  {Jiang}}, \bibinfo {author} {\bibfnamefont {X.}~\bibnamefont {Qian}},
  \bibinfo {author} {\bibfnamefont {R.}~\bibnamefont {Yang}}, \ and\ \bibinfo
  {author} {\bibfnamefont {L.}~\bibnamefont {Lindsay}},\ }\href@noop {}
  {\bibfield  {journal} {\bibinfo  {journal} {Physical Review Materials}\
  }\textbf {\bibinfo {volume} {2}},\ \bibinfo {pages} {64005} (\bibinfo {year}
  {2018}{\natexlab{a}})}\BibitemShut {NoStop}%
\bibitem [{\citenamefont {Kubi{\'{s}}}\ \emph {et~al.}(2020)\citenamefont
  {Kubi{\'{s}}}, \citenamefont {Pietrak}, \citenamefont {Cie{\'{s}}likiewicz},
  \citenamefont {Furma{\'{n}}ski}, \citenamefont {Wasik}, \citenamefont
  {Seredy{\'{n}}ski}, \citenamefont {Wi{\'{s}}niewski},\ and\ \citenamefont
  {{\L}apka}}]{Kubis2020}%
  \BibitemOpen
  \bibfield  {author} {\bibinfo {author} {\bibfnamefont {M.}~\bibnamefont
  {Kubi{\'{s}}}}, \bibinfo {author} {\bibfnamefont {K.}~\bibnamefont
  {Pietrak}}, \bibinfo {author} {\bibfnamefont {{\L}.}~\bibnamefont
  {Cie{\'{s}}likiewicz}}, \bibinfo {author} {\bibfnamefont {P.}~\bibnamefont
  {Furma{\'{n}}ski}}, \bibinfo {author} {\bibfnamefont {M.}~\bibnamefont
  {Wasik}}, \bibinfo {author} {\bibfnamefont {M.}~\bibnamefont
  {Seredy{\'{n}}ski}}, \bibinfo {author} {\bibfnamefont {T.~S.}\ \bibnamefont
  {Wi{\'{s}}niewski}}, \ and\ \bibinfo {author} {\bibfnamefont
  {P.}~\bibnamefont {{\L}apka}},\ }\href@noop {} {\bibfield  {journal}
  {\bibinfo  {journal} {Journal of Building Engineering}\ }\textbf {\bibinfo
  {volume} {31}} (\bibinfo {year} {2020})}\BibitemShut {NoStop}%
\bibitem [{\citenamefont {Sun}\ \emph {et~al.}(2019)\citenamefont {Sun},
  \citenamefont {Haunschild}, \citenamefont {Polanco}, \citenamefont {Ju},
  \citenamefont {Lindsay}, \citenamefont {Koblm{\"{u}}ller},\ and\
  \citenamefont {Koh}}]{Sun2019}%
  \BibitemOpen
  \bibfield  {author} {\bibinfo {author} {\bibfnamefont {B.}~\bibnamefont
  {Sun}}, \bibinfo {author} {\bibfnamefont {G.}~\bibnamefont {Haunschild}},
  \bibinfo {author} {\bibfnamefont {C.}~\bibnamefont {Polanco}}, \bibinfo
  {author} {\bibfnamefont {J.~Z.~J.}\ \bibnamefont {Ju}}, \bibinfo {author}
  {\bibfnamefont {L.}~\bibnamefont {Lindsay}}, \bibinfo {author} {\bibfnamefont
  {G.}~\bibnamefont {Koblm{\"{u}}ller}}, \ and\ \bibinfo {author}
  {\bibfnamefont {Y.~K.}\ \bibnamefont {Koh}},\ }\href@noop {} {\bibfield
  {journal} {\bibinfo  {journal} {Nature Materials}\ }\textbf {\bibinfo
  {volume} {18}},\ \bibinfo {pages} {136} (\bibinfo {year} {2019})}\BibitemShut
  {NoStop}%
\bibitem [{\citenamefont {Ly}\ \emph {et~al.}(2021)\citenamefont {Ly},
  \citenamefont {Ngoc}, \citenamefont {Kang},\ and\ \citenamefont
  {Choi}}]{Ly2021}%
  \BibitemOpen
  \bibfield  {author} {\bibinfo {author} {\bibfnamefont {L.}~\bibnamefont
  {Ly}}, \bibinfo {author} {\bibfnamefont {P.}~\bibnamefont {Ngoc}}, \bibinfo
  {author} {\bibfnamefont {K.}~\bibnamefont {Kang}}, \ and\ \bibinfo {author}
  {\bibfnamefont {G.-M.}\ \bibnamefont {Choi}},\ }\href@noop {} {\bibfield
  {journal} {\bibinfo  {journal} {AIP Advances}\ }\textbf {\bibinfo {volume}
  {11}},\ \bibinfo {pages} {25024} (\bibinfo {year} {2021})}\BibitemShut
  {NoStop}%
\bibitem [{\citenamefont {Zhu}\ \emph {et~al.}(2014)\citenamefont {Zhu},
  \citenamefont {Hippalgaonkar}, \citenamefont {Shen}, \citenamefont {Wang},
  \citenamefont {Abate}, \citenamefont {Lee}, \citenamefont {Wu}, \citenamefont
  {Yin}, \citenamefont {Majumdar},\ and\ \citenamefont {Zhang}}]{Zhu2014}%
  \BibitemOpen
  \bibfield  {author} {\bibinfo {author} {\bibfnamefont {J.}~\bibnamefont
  {Zhu}}, \bibinfo {author} {\bibfnamefont {K.}~\bibnamefont {Hippalgaonkar}},
  \bibinfo {author} {\bibfnamefont {S.}~\bibnamefont {Shen}}, \bibinfo {author}
  {\bibfnamefont {K.}~\bibnamefont {Wang}}, \bibinfo {author} {\bibfnamefont
  {Y.}~\bibnamefont {Abate}}, \bibinfo {author} {\bibfnamefont
  {S.}~\bibnamefont {Lee}}, \bibinfo {author} {\bibfnamefont {J.}~\bibnamefont
  {Wu}}, \bibinfo {author} {\bibfnamefont {X.}~\bibnamefont {Yin}}, \bibinfo
  {author} {\bibfnamefont {A.}~\bibnamefont {Majumdar}}, \ and\ \bibinfo
  {author} {\bibfnamefont {X.}~\bibnamefont {Zhang}},\ }\href@noop {}
  {\bibfield  {journal} {\bibinfo  {journal} {Nano Letters}\ }\textbf {\bibinfo
  {volume} {14}},\ \bibinfo {pages} {4867} (\bibinfo {year}
  {2014})}\BibitemShut {NoStop}%
\bibitem [{\citenamefont {Li}\ \emph {et~al.}(2012)\citenamefont {Li},
  \citenamefont {Ren}, \citenamefont {Wang}, \citenamefont {Zhang},
  \citenamefont {H{\"{a}}nggi},\ and\ \citenamefont {Li}}]{Li2012h}%
  \BibitemOpen
  \bibfield  {author} {\bibinfo {author} {\bibfnamefont {N.}~\bibnamefont
  {Li}}, \bibinfo {author} {\bibfnamefont {J.}~\bibnamefont {Ren}}, \bibinfo
  {author} {\bibfnamefont {L.}~\bibnamefont {Wang}}, \bibinfo {author}
  {\bibfnamefont {G.}~\bibnamefont {Zhang}}, \bibinfo {author} {\bibfnamefont
  {P.}~\bibnamefont {H{\"{a}}nggi}}, \ and\ \bibinfo {author} {\bibfnamefont
  {B.}~\bibnamefont {Li}},\ }\href@noop {} {\bibfield  {journal} {\bibinfo
  {journal} {Reviews of Modern Physics}\ }\textbf {\bibinfo {volume} {84}},\
  \bibinfo {pages} {1045} (\bibinfo {year} {2012})}\BibitemShut {NoStop}%
\bibitem [{\citenamefont {Maldovan}(2013)}]{Maldovan2013b}%
  \BibitemOpen
  \bibfield  {author} {\bibinfo {author} {\bibfnamefont {M.}~\bibnamefont
  {Maldovan}},\ }\href@noop {} {\bibfield  {journal} {\bibinfo  {journal}
  {Nautre}\ }\textbf {\bibinfo {volume} {503}},\ \bibinfo {pages} {209}
  (\bibinfo {year} {2013})}\BibitemShut {NoStop}%
\bibitem [{\citenamefont {Li}\ \emph {et~al.}(2021)\citenamefont {Li},
  \citenamefont {Li}, \citenamefont {Han}, \citenamefont {Zheng}, \citenamefont
  {Li}, \citenamefont {Li}, \citenamefont {Fan},\ and\ \citenamefont
  {Qiu}}]{Li2021}%
  \BibitemOpen
  \bibfield  {author} {\bibinfo {author} {\bibfnamefont {Y.}~\bibnamefont
  {Li}}, \bibinfo {author} {\bibfnamefont {W.}~\bibnamefont {Li}}, \bibinfo
  {author} {\bibfnamefont {T.}~\bibnamefont {Han}}, \bibinfo {author}
  {\bibfnamefont {X.}~\bibnamefont {Zheng}}, \bibinfo {author} {\bibfnamefont
  {J.}~\bibnamefont {Li}}, \bibinfo {author} {\bibfnamefont {B.}~\bibnamefont
  {Li}}, \bibinfo {author} {\bibfnamefont {S.}~\bibnamefont {Fan}}, \ and\
  \bibinfo {author} {\bibfnamefont {C.~W.}\ \bibnamefont {Qiu}},\ }\href@noop
  {} {\bibfield  {journal} {\bibinfo  {journal} {Nature Reviews Materials}\
  }\textbf {\bibinfo {volume} {6}},\ \bibinfo {pages} {488} (\bibinfo {year}
  {2021})}\BibitemShut {NoStop}%
\bibitem [{\citenamefont {Tong}\ and\ \citenamefont
  {Majumdar}(2006)}]{Tong2006}%
  \BibitemOpen
  \bibfield  {author} {\bibinfo {author} {\bibfnamefont {T.}~\bibnamefont
  {Tong}}\ and\ \bibinfo {author} {\bibfnamefont {A.}~\bibnamefont
  {Majumdar}},\ }\href@noop {} {\bibfield  {journal} {\bibinfo  {journal}
  {Review of Scientific Instruments}\ }\textbf {\bibinfo {volume} {77}},\
  \bibinfo {pages} {104902} (\bibinfo {year} {2006})}\BibitemShut {NoStop}%
\bibitem [{\citenamefont {Ramu}\ and\ \citenamefont {Bowers}(2012)}]{Ramu2012}%
  \BibitemOpen
  \bibfield  {author} {\bibinfo {author} {\bibfnamefont {A.~T.}\ \bibnamefont
  {Ramu}}\ and\ \bibinfo {author} {\bibfnamefont {J.~E.}\ \bibnamefont
  {Bowers}},\ }\href@noop {} {\bibfield  {journal} {\bibinfo  {journal} {Review
  of Scientific Instruments}\ }\textbf {\bibinfo {volume} {83}},\ \bibinfo
  {pages} {124903} (\bibinfo {year} {2012})}\BibitemShut {NoStop}%
\bibitem [{\citenamefont {Mishra}\ \emph {et~al.}(2015)\citenamefont {Mishra},
  \citenamefont {Hardin}, \citenamefont {Garay},\ and\ \citenamefont
  {Dames}}]{Mishra2015a}%
  \BibitemOpen
  \bibfield  {author} {\bibinfo {author} {\bibfnamefont {V.}~\bibnamefont
  {Mishra}}, \bibinfo {author} {\bibfnamefont {C.~L.}\ \bibnamefont {Hardin}},
  \bibinfo {author} {\bibfnamefont {J.~E.}\ \bibnamefont {Garay}}, \ and\
  \bibinfo {author} {\bibfnamefont {C.}~\bibnamefont {Dames}},\ }\href@noop {}
  {\bibfield  {journal} {\bibinfo  {journal} {Review of Scientific
  Instruments}\ }\textbf {\bibinfo {volume} {86}},\ \bibinfo {pages} {054902}
  (\bibinfo {year} {2015})}\BibitemShut {NoStop}%
\bibitem [{\citenamefont {Schmidt}\ \emph {et~al.}(2008)\citenamefont
  {Schmidt}, \citenamefont {Chen},\ and\ \citenamefont {Chen}}]{Schmidt2008}%
  \BibitemOpen
  \bibfield  {author} {\bibinfo {author} {\bibfnamefont {A.~J.}\ \bibnamefont
  {Schmidt}}, \bibinfo {author} {\bibfnamefont {X.}~\bibnamefont {Chen}}, \
  and\ \bibinfo {author} {\bibfnamefont {G.}~\bibnamefont {Chen}},\ }\href@noop
  {} {\bibfield  {journal} {\bibinfo  {journal} {Review of Scientific
  Instruments}\ }\textbf {\bibinfo {volume} {79}},\ \bibinfo {pages} {114902}
  (\bibinfo {year} {2008})}\BibitemShut {NoStop}%
\bibitem [{\citenamefont {Feser}\ and\ \citenamefont
  {Cahill}(2012)}]{Feser2012a}%
  \BibitemOpen
  \bibfield  {author} {\bibinfo {author} {\bibfnamefont {J.~P.}\ \bibnamefont
  {Feser}}\ and\ \bibinfo {author} {\bibfnamefont {D.~G.}\ \bibnamefont
  {Cahill}},\ }\href@noop {} {\bibfield  {journal} {\bibinfo  {journal} {Review
  of Scientific Instruments}\ }\textbf {\bibinfo {volume} {83}},\ \bibinfo
  {pages} {104901} (\bibinfo {year} {2012})}\BibitemShut {NoStop}%
\bibitem [{\citenamefont {Feser}\ \emph {et~al.}(2014)\citenamefont {Feser},
  \citenamefont {Liu},\ and\ \citenamefont {Cahill}}]{Feser2014}%
  \BibitemOpen
  \bibfield  {author} {\bibinfo {author} {\bibfnamefont {J.~P.}\ \bibnamefont
  {Feser}}, \bibinfo {author} {\bibfnamefont {J.}~\bibnamefont {Liu}}, \ and\
  \bibinfo {author} {\bibfnamefont {D.~G.}\ \bibnamefont {Cahill}},\
  }\href@noop {} {\bibfield  {journal} {\bibinfo  {journal} {Review of
  Scientific Instruments}\ }\textbf {\bibinfo {volume} {85}},\ \bibinfo {pages}
  {104903} (\bibinfo {year} {2014})}\BibitemShut {NoStop}%
\bibitem [{\citenamefont {Rodin}\ and\ \citenamefont {Yee}(2017)}]{Rodin2017}%
  \BibitemOpen
  \bibfield  {author} {\bibinfo {author} {\bibfnamefont {D.}~\bibnamefont
  {Rodin}}\ and\ \bibinfo {author} {\bibfnamefont {S.~K.}\ \bibnamefont
  {Yee}},\ }\href@noop {} {\bibfield  {journal} {\bibinfo  {journal} {Review of
  Scientific Instruments}\ }\textbf {\bibinfo {volume} {88}},\ \bibinfo {pages}
  {014902} (\bibinfo {year} {2017})}\BibitemShut {NoStop}%
\bibitem [{\citenamefont {Jiang}\ \emph {et~al.}(2017)\citenamefont {Jiang},
  \citenamefont {Qian},\ and\ \citenamefont {Yang}}]{Jiang2017}%
  \BibitemOpen
  \bibfield  {author} {\bibinfo {author} {\bibfnamefont {P.}~\bibnamefont
  {Jiang}}, \bibinfo {author} {\bibfnamefont {X.}~\bibnamefont {Qian}}, \ and\
  \bibinfo {author} {\bibfnamefont {R.}~\bibnamefont {Yang}},\ }\href@noop {}
  {\bibfield  {journal} {\bibinfo  {journal} {Review of Scientific
  Instruments}\ }\textbf {\bibinfo {volume} {88}},\ \bibinfo {pages} {074901}
  (\bibinfo {year} {2017})}\BibitemShut {NoStop}%
\bibitem [{\citenamefont {Jiang}\ \emph
  {et~al.}(2018{\natexlab{b}})\citenamefont {Jiang}, \citenamefont {Qian},\
  and\ \citenamefont {Yang}}]{Jiang2018}%
  \BibitemOpen
  \bibfield  {author} {\bibinfo {author} {\bibfnamefont {P.}~\bibnamefont
  {Jiang}}, \bibinfo {author} {\bibfnamefont {X.}~\bibnamefont {Qian}}, \ and\
  \bibinfo {author} {\bibfnamefont {R.}~\bibnamefont {Yang}},\ }\href@noop {}
  {\bibfield  {journal} {\bibinfo  {journal} {Review of Scientific
  Instruments}\ }\textbf {\bibinfo {volume} {89}},\ \bibinfo {pages} {094902}
  (\bibinfo {year} {2018}{\natexlab{b}})}\BibitemShut {NoStop}%
\bibitem [{\citenamefont {Li}\ \emph {et~al.}(2018)\citenamefont {Li},
  \citenamefont {Kang},\ and\ \citenamefont {Hu}}]{Li2018}%
  \BibitemOpen
  \bibfield  {author} {\bibinfo {author} {\bibfnamefont {M.}~\bibnamefont
  {Li}}, \bibinfo {author} {\bibfnamefont {J.~S.}\ \bibnamefont {Kang}}, \ and\
  \bibinfo {author} {\bibfnamefont {Y.}~\bibnamefont {Hu}},\ }\href@noop {}
  {\bibfield  {journal} {\bibinfo  {journal} {Review of Scientific
  Instruments}\ }\textbf {\bibinfo {volume} {89}},\ \bibinfo {pages} {084901}
  (\bibinfo {year} {2018})}\BibitemShut {NoStop}%
\bibitem [{\citenamefont {Rahman}\ \emph {et~al.}(2018)\citenamefont {Rahman},
  \citenamefont {Shahzadeh}, \citenamefont {Braeuninger-Weimer}, \citenamefont
  {Hofmann}, \citenamefont {Hellwig},\ and\ \citenamefont
  {Pisana}}]{Rahman2018}%
  \BibitemOpen
  \bibfield  {author} {\bibinfo {author} {\bibfnamefont {M.}~\bibnamefont
  {Rahman}}, \bibinfo {author} {\bibfnamefont {M.}~\bibnamefont {Shahzadeh}},
  \bibinfo {author} {\bibfnamefont {P.}~\bibnamefont {Braeuninger-Weimer}},
  \bibinfo {author} {\bibfnamefont {S.}~\bibnamefont {Hofmann}}, \bibinfo
  {author} {\bibfnamefont {O.}~\bibnamefont {Hellwig}}, \ and\ \bibinfo
  {author} {\bibfnamefont {S.}~\bibnamefont {Pisana}},\ }\href@noop {}
  {\bibfield  {journal} {\bibinfo  {journal} {Journal of Applied Physics}\
  }\textbf {\bibinfo {volume} {123}} (\bibinfo {year} {2018})}\BibitemShut
  {NoStop}%
\bibitem [{\citenamefont {Yuan}\ \emph {et~al.}(2019)\citenamefont {Yuan},
  \citenamefont {Waller},\ and\ \citenamefont {Kuball}}]{Yuan2019}%
  \BibitemOpen
  \bibfield  {author} {\bibinfo {author} {\bibfnamefont {C.}~\bibnamefont
  {Yuan}}, \bibinfo {author} {\bibfnamefont {W.~M.}\ \bibnamefont {Waller}}, \
  and\ \bibinfo {author} {\bibfnamefont {M.}~\bibnamefont {Kuball}},\
  }\href@noop {} {\bibfield  {journal} {\bibinfo  {journal} {Rev. Sci.
  Instrum}\ }\textbf {\bibinfo {volume} {90}},\ \bibinfo {pages} {114903}
  (\bibinfo {year} {2019})}\BibitemShut {NoStop}%
\bibitem [{\citenamefont {Qian}\ \emph {et~al.}(2020)\citenamefont {Qian},
  \citenamefont {Ding}, \citenamefont {Shin}, \citenamefont {Schmidt},\ and\
  \citenamefont {Chen}}]{Qian2020}%
  \BibitemOpen
  \bibfield  {author} {\bibinfo {author} {\bibfnamefont {X.}~\bibnamefont
  {Qian}}, \bibinfo {author} {\bibfnamefont {Z.}~\bibnamefont {Ding}}, \bibinfo
  {author} {\bibfnamefont {J.}~\bibnamefont {Shin}}, \bibinfo {author}
  {\bibfnamefont {A.~J.}\ \bibnamefont {Schmidt}}, \ and\ \bibinfo {author}
  {\bibfnamefont {G.}~\bibnamefont {Chen}},\ }\href@noop {} {\bibfield
  {journal} {\bibinfo  {journal} {Review of Scientific Instruments}\ }\textbf
  {\bibinfo {volume} {91}},\ \bibinfo {pages} {064903} (\bibinfo {year}
  {2020})}\BibitemShut {NoStop}%
\bibitem [{\citenamefont {Tang}\ and\ \citenamefont {Dames}(2021)}]{Tang2021}%
  \BibitemOpen
  \bibfield  {author} {\bibinfo {author} {\bibfnamefont {L.}~\bibnamefont
  {Tang}}\ and\ \bibinfo {author} {\bibfnamefont {C.}~\bibnamefont {Dames}},\
  }\href@noop {} {\bibfield  {journal} {\bibinfo  {journal} {International
  Journal of Heat and Mass Transfer}\ }\textbf {\bibinfo {volume} {164}},\
  \bibinfo {pages} {120600} (\bibinfo {year} {2021})}\BibitemShut {NoStop}%
\bibitem [{\citenamefont {Cahill}(1990)}]{Cahill1990}%
  \BibitemOpen
  \bibfield  {author} {\bibinfo {author} {\bibfnamefont {D.~G.}\ \bibnamefont
  {Cahill}},\ }\href@noop {} {\bibfield  {journal} {\bibinfo  {journal} {Review
  of Scientific Instruments}\ }\textbf {\bibinfo {volume} {61}},\ \bibinfo
  {pages} {802} (\bibinfo {year} {1990})}\BibitemShut {NoStop}%
\bibitem [{\citenamefont {Cahill}(2004)}]{Cahill2004}%
  \BibitemOpen
  \bibfield  {author} {\bibinfo {author} {\bibfnamefont {D.~G.}\ \bibnamefont
  {Cahill}},\ }\href@noop {} {\bibfield  {journal} {\bibinfo  {journal} {Review
  of Scientific Instruments}\ }\textbf {\bibinfo {volume} {75}},\ \bibinfo
  {pages} {5119} (\bibinfo {year} {2004})}\BibitemShut {NoStop}%
\bibitem [{\citenamefont {Kwon}\ \emph {et~al.}(2017)\citenamefont {Kwon},
  \citenamefont {Zheng}, \citenamefont {Wingert}, \citenamefont {Cui},\ and\
  \citenamefont {Chen}}]{Kwon2017}%
  \BibitemOpen
  \bibfield  {author} {\bibinfo {author} {\bibfnamefont {S.}~\bibnamefont
  {Kwon}}, \bibinfo {author} {\bibfnamefont {J.}~\bibnamefont {Zheng}},
  \bibinfo {author} {\bibfnamefont {M.~C.}\ \bibnamefont {Wingert}}, \bibinfo
  {author} {\bibfnamefont {S.}~\bibnamefont {Cui}}, \ and\ \bibinfo {author}
  {\bibfnamefont {R.}~\bibnamefont {Chen}},\ }\href@noop {} {\  (\bibinfo
  {year} {2017})}\BibitemShut {NoStop}%
\bibitem [{\citenamefont {Borca-Tasciuc}\ \emph {et~al.}(2001)\citenamefont
  {Borca-Tasciuc}, \citenamefont {Kumar}, \citenamefont {Chen},\ and\
  \citenamefont {Kumar}}]{Borca-Tasciuc2001}%
  \BibitemOpen
  \bibfield  {author} {\bibinfo {author} {\bibfnamefont {T.}~\bibnamefont
  {Borca-Tasciuc}}, \bibinfo {author} {\bibfnamefont {A.~R.}\ \bibnamefont
  {Kumar}}, \bibinfo {author} {\bibfnamefont {G.}~\bibnamefont {Chen}}, \ and\
  \bibinfo {author} {\bibfnamefont {.~A.~R.}\ \bibnamefont {Kumar}},\
  }\href@noop {} {\bibfield  {journal} {\bibinfo  {journal} {Review of
  Scientific Instruments}\ }\textbf {\bibinfo {volume} {72}},\ \bibinfo {pages}
  {2590} (\bibinfo {year} {2001})}\BibitemShut {NoStop}%
\bibitem [{\citenamefont {Carslaw}\ and\ \citenamefont
  {Jaeger}(1986)}]{Carslaw1986}%
  \BibitemOpen
  \bibfield  {author} {\bibinfo {author} {\bibfnamefont {H.~S.}\ \bibnamefont
  {Carslaw}}\ and\ \bibinfo {author} {\bibfnamefont {J.~J.~C.}\ \bibnamefont
  {Jaeger}},\ }\href@noop {} {\emph {\bibinfo {title} {{"Conduction of Heat in
  Solids"}}}}\ (\bibinfo  {publisher} {Oxford University Press},\ \bibinfo
  {year} {1986})\ p.\ \bibinfo {pages} {510}\BibitemShut {NoStop}%
\bibitem [{\citenamefont {Favaloro}\ \emph {et~al.}(2015)\citenamefont
  {Favaloro}, \citenamefont {Bahk},\ and\ \citenamefont
  {Shakouri}}]{Favaloro2015}%
  \BibitemOpen
  \bibfield  {author} {\bibinfo {author} {\bibfnamefont {T.}~\bibnamefont
  {Favaloro}}, \bibinfo {author} {\bibfnamefont {J.~H.}\ \bibnamefont {Bahk}},
  \ and\ \bibinfo {author} {\bibfnamefont {A.}~\bibnamefont {Shakouri}},\
  }\href@noop {} {\bibfield  {journal} {\bibinfo  {journal} {Review of
  Scientific Instruments}\ }\textbf {\bibinfo {volume} {86}},\ \bibinfo {pages}
  {024903} (\bibinfo {year} {2015})}\BibitemShut {NoStop}%
\end{thebibliography}%
\end{document}